  \documentclass[a4paper, 12pt, pre, nofootinbib, onecolumn, groupedaddress, floatfix]{revtex4}
  \usepackage{amsmath, amstext, amssymb, amsfonts,amsmath, mathtools}
  \usepackage{url,microtype}
  \usepackage[dvips]{graphicx}
  \usepackage{exscale}
  \usepackage{subfigure}
  \usepackage[dvips, breaklinks, colorlinks=true,urlcolor=blue,citecolor=blue,linkcolor=blue,hyperindex]{hyperref}
  \usepackage[a4paper, top=2.3cm, bottom=1.1cm, right=1.5cm, left=1.5cm]{geometry}
  \usepackage{breakurl}
 \graphicspath{ {./FIG/} }
 \newcommand{\rd}{\mathrm{d}}
 \newcommand{\Ha}{\mathcal{H}}
 \newcommand{\de}[1]{\left(#1\right)}
 \newcommand{\comu}[1]{\left[#1\right]}
 \newcommand{\Ntil}{\widetilde{N}}
 \newcommand{\df}[2]{\dfrac{#1}{#2}}
 \newcommand{\SBG}{S_{\mathrm{BG}}}
 \newcommand{\media}[1]{\left\langle #1 \right\rangle}
 \let \pa  = \partial
 \renewcommand{\baselinestretch}{1.25}
 
\begin{document}
 \renewcommand{\baselinestretch}{1.10}
 \title{Validity and Failure of the Boltzmann Weight}
 \author{Leonardo J.\ L.\ Cirto$^{1}$}    \email{cirto@cbpf.br}
 \author{Antonio Rodr\'iguez$^{2}$}       \email{antonio.rodriguezm@upm.es}
 \author{Fernando D.\ Nobre$^{1,3}$}      \email{fdnobre@cbpf.br}
 \author{Constantino Tsallis$^{1,3,4,5}$} \email{tsallis@cbpf.br}
 \affiliation{$^{1}$\footnotesize{Centro Brasileiro de Pesquisas F\'{\i}sicas, Rua Dr.\ Xavier Sigaud 150, 22290-180 Rio de Janeiro, Brazil}}
 \affiliation{$^{2}$\footnotesize{Departamento de Matem\'atica Aplicada a la Ingenier\'ia Aeroespacial, Universidad Polit\'ecnica de Madrid, Plaza Cardenal Cisneros s/n, 28040 Madrid, Spain}}
 \affiliation{$^{3}$\footnotesize{National Institute of Science and Technology for Complex Systems, Rua Dr.\ Xavier Sigaud 150, 22290-180 Rio de Janeiro, Brazil}}
 \affiliation{$^{4}$\footnotesize{Santa Fe Institute, 1399 Hyde Park Road, Santa Fe, 87501 New Mexico, United States}}
 \affiliation{$^{5}$\footnotesize{Complexity Science Hub Vienna, Josefst\"adter Strasse 39, 1080 Vienna, Austria}}
\begin{abstract}
The dynamics and thermostatistics of a classical inertial XY model, characterized by long-range interactions, are investigated on $d$-dimensional lattices ($d=1,2,$ and 3), through molecular dynamics.
The interactions between rotators decay with the distance $r_{ij}$ like~$1/r_{ij}^{\alpha}$ ($\alpha \geq 0$), where $\alpha\to\infty$ and $\alpha=0$ respectively correspond to the nearest-neighbor and infinite-range interactions.
We verify that the momenta probability distributions are Maxwellians in the short-range regime, whereas $q$-Gaussians emerge in the long-range regime.
Moreover, in this latter regime, the individual energy probability distributions are characterized by long tails, corresponding to $q$-exponential functions.
The present investigation strongly indicates that, in the long-range regime, central properties fall out of the scope of Boltzmann-Gibbs statistical mechanics, depending on $d$ and $\alpha$  through the ratio $\alpha/d$.
\end{abstract}
 \maketitle
Systems with long-range-interacting elements have been object of many researches and controversies.
Usually, these systems are addressed through statistical-mechanical
techniques, and they cover from physical, biological, and mathematical models to complex 
networks~\cite{CampaDauxoisRuffoPR2009,Lynden_BellMNRAS1967Lynden_BellWoodMNRAS1968,JundKimTsallisPRB1995,TsallisFractals1995,
AnteneodoTsallisPRL1998,
TamaritAnteneodoPRL2000,AntoniRuffoPRE1995,LuijtenBlotePRL2002,
VegaSanchezNPB2002,
CampaGiansantiMoroniJPA2003,%
PluchinoRapisardaTsallisEPL2007,PluchinoRapisardaTsallisPA2008,%
BachelardKastnerPRL2013,%
ChristodoulidiTsallisBountisEPL2014,CirtoAssisTsallisPA2014,%
CirtoLimaNobreJSTAT2015,%
BagchiTsallisPRE2016,BritoSilvaTsallisSR2016,NunesBritoSilvaTsallisJSTAT2017,
ChristodoulidiBountisTsallisDrossosJSTAT2016,GuptaRuffoIJMPA2017}.
Interesting phenomena, like breakdown of ergodicity, 
nonequivalence of statistical ensembles, and long-lived quasistationary states,
emerge frequently when long-range forces 
come into play; such situations usually fall out of the scope of  
Boltzmann-Gibbs (BG) statistical mechanics, which has been developed 
assuming explicitly, or tacitly, short-range interactions between elements.

For $N$-body Hamiltonian systems in $d$-dimensions, with a potential~$\Phi(r)\propto1/r^{\alpha}$,
the case $\alpha/d=1$ represents a threshold between long- and short-range regimes.
By considering {a model ruled by} this power-law dependence, 
one interpolates
between the infinite-range-interaction ($\alpha=0$) and the nearest-neighbor
($\alpha\to\infty$) limits, allowing to investigate the influence
of the interaction range on the thermostatistics of the model.
In the long-range regime, corresponding to $0\leq\alpha\leq d$, the 
potential energy, as well as the total energy, scale superlinearly 
with~$N$ so that the system is said to be nonextensive. 
In this case, BG statistical mechanics faces several difficulties.
In order to derive thermodynamic properties 
one may redefine the thermodynamic 
limit~\cite{JundKimTsallisPRB1995,TsallisFractals1995,%
AnteneodoTsallisPRL1998,VegaSanchezNPB2002},
or employ a properly generalized Kac's prescription, by 
weakening the strength of interparticle forces as the system size~$N$ 
increases -- an artificial modification of the model to turn 
its total energy extensive.
However, even if extensivity is formally recovered by a microscopic modification
like Kac's prescription, the system preserves its long-range nature for small values of $\alpha$. 
For instance, one may mention the behavior of the Lyapunov 
exponent~\cite{AnteneodoTsallisPRL1998, BagchiTsallisPRE2016}, 
the presence of non-Boltzmannian quasistationary
states~{(QSSs)}~\cite{EttoumiFirpoPRE2013, BachelardKastnerPRL2013, CirtoAssisTsallisPA2014,%
CirtoLimaNobreJSTAT2015}, the emergence of typical nonextensive features like $q$-Gaussians and
$q$-exponentials~\cite{CirtoAssisTsallisPA2014,ChristodoulidiTsallisBountisEPL2014, ChristodoulidiBountisTsallisDrossosJSTAT2016}, among others. 
Some of those features remain beyond the threshold~$\alpha/d=1$, where the system can still 
display long-range properties~(see, e.g., Ref.~\cite{LuijtenBlotePRL2002} 
and references therein).

A paradigmatic Hamiltonian model, frequently used to investigate several of the above-mentioned properties, is defined in terms of classical XY rotators
with infinite-range interactions. In this limit, the  BG equilibrium state is exactly tractable within the standard mean-field approach.
Due to this, this model is usually known in the literature as Hamiltonian-Mean-Field (HMF) one~\cite{AntoniRuffoPRE1995, CampaDauxoisRuffoPR2009}.
The HMF model has been numerically and analytically studied intensively in the last two decades, and one of its intriguing features concerns the existence of QSSs whose lifetime diverges as the system size~$N$ increases.
Particularly, these QSSs exhibit a clear breakdown of 
ergodicity, in the sense that velocity distributions calculated from time averages~\cite{PluchinoRapisardaTsallisEPL2007,PluchinoRapisardaTsallisPA2008} 
appear to be quite different from those of ensemble 
averages~\cite{CampaDauxoisRuffoPR2009, CirtoAssisTsallisPA2014, PluchinoRapisardaTsallisEPL2007,PluchinoRapisardaTsallisPA2008,%
AntoniazziFanelliBarreChavanisDauxoisRuffoPRE2007,AntoniazziCalifanoFanelliRuffoPRL2007,%
AntoniazziFanelliRuffoYamaguchiPRL2007,MartelloniMartelloniBuylFanelliPRE2016}.
Moreover, for any finite $N$, the QSSs are followed by a second plateau at longer times, which presents a kinetic temperature that coincides
with the one calculated analytically from BG statistical mechanics, although it also exhibits further 
curious properties, like time-averaged long-tailed velocity distributions~\cite{CirtoAssisTsallisPA2014}, in notorious contrast with the BG theory. 

Herein we present a numerical analysis of the so-called $\alpha$-XY model, 
which consists of a classical~XY rotator  
system, with controllable range of (ferromagnetic) interactions,
decaying like~$1/r^{\alpha}$. 
Previous results ($\alpha=0$ and~$d=1$) revealed, among other non-standard features,  
non-Maxwellian velocity distributions~\cite{PluchinoRapisardaTsallisEPL2007, PluchinoRapisardaTsallisPA2008,%
CirtoAssisTsallisPA2014}. 
In the short-range regime, on the other hand, all the standard 
BG results are recovered, including, naturally, the Maxwellian distribution. 
It has been verified that these non-Maxwellian distributions are well fitted along several decades with the so-called $q$-Gaussians, 
landmark functions of nonextensive statistical mechanics, built on the basis of the nonadditive 
entropy~$S_q$~\cite{TsallisJSP1988, TsallisMendesPlastinoPA1998, TsallisLivro2009}; similarly,  Maxwellians represent a landmark of BG statistics, 
built from the additive entropy~$\SBG$.
Indeed, the entropy $S_q$ is defined as a generalization of~$\SBG$,  
\begin{equation}
S_q = k \sum_{i=1}^W p_i \ln_q \dfrac{1}{p_i}
\hspace{0.5cm}\comu{\ln_q x \equiv \dfrac{x^{1-q}-1}{1-q}; \ (x>0) }, 
\end{equation}
where $W$ accounts for the microscopic configurations, and $\ln_q x$ denotes the {\it $q$-logarithm}; we verify 
that $\SBG\equiv -k \sum p_i\ln p_i=\lim_{q\to1}S_q$. 
The {\it $q$-exponential}~$\exp_q[-\beta x]=[1-\beta (1-q)x]^{1/(1-q)}$ and 
the {\it $q$-Gaussian}~$\exp_q[-\beta x^2]$ appear naturally by extremizing 
$S_q$ under appropriate 
constraints~\cite{TsallisJSP1988, TsallisMendesPlastinoPA1998, TsallisLivro2009}.  
The generalized thermostatistics based on~$S_q$ frequently applies when
assumptions underlying the BG thermostatistics are not fulfilled
(like, e.g., mixing and ergodicity)~\cite{TsallisLivro2009, CirtoAssisTsallisPA2014, BritoSilvaTsallisSR2016,NunesBritoSilvaTsallisJSTAT2017,ChristodoulidiTsallisBountisEPL2014, ChristodoulidiBountisTsallisDrossosJSTAT2016, WongWilkCirtoTsallisPRD2015, YalcinBeckSR2018,  BetzlerBorgesBetzlerBorgesMNRAS2015, TirnakliBorgesSR2016, CombeRichefeuStasiakAtmanPRL2015}.

In the present work we are primarily interested on how $\alpha/d$ influences 
the one-particle velocity and energy distributions of the $\alpha$-XY 
{model, focusing mainly on the second plateau that follows the QSS at longer times.} 
We explore how higher values of~$d$ modify previous~$d=1$ results for the 
velocity distribution; moreover, we show how the energy distribution changes from the celebrated
exponential Boltzmann weight 
to a distribution well described by a $q$-exponential, as the system goes from the short- 
to the long-range regime.
We have also verified a universal scaling law of these distributions
governed by the~$\alpha/d$ ratio, similarly to what was found in other 
complex systems~\cite{BritoSilvaTsallisSR2016,NunesBritoSilvaTsallisJSTAT2017}.

We consider the $\alpha$-XY model on 
$d$-dimensional hypercubic lattices, defined by a Hamiltonian,
$\Ha=K + V$ (kinetic and potential contributions, respectively), 
conveniently written below in terms of one-particle energies $E_i$, 
\begin{equation}
\Ha = \sum_{\substack{i=1}}^{\substack{N}} E_i~;
\hspace{0.5cm}
E_i = \frac{1}{2} p_i^2 + \frac{1}{2\Ntil}\sum_{\substack{j\neq i}}^{\substack{N}} \frac{1- \cos\de{\theta_i-\theta_j} }{r_{ij}^\alpha}  \,.
\label{eq:Hamiltonian_XY}
\end{equation}

At a given time $t$, each rotator $i$ ($i=1,2, \dots , N$) is characterized by the 
angle~$\theta_i(t)$ and its conjugated momentum~$p_i(t)$,
so that the dynamics of the system follows from the Hamilton equations of motion, 
\begin{equation}
\dot{\theta}_i = \frac{\pa \Ha}{\pa p_i} = p_i;
\hspace{0.5cm}
\dot{p}_i = -\frac{\pa \Ha}{\pa \theta_i} 
=-\frac{1}{\Ntil}\sum_{\substack{j\neq i}}^{\substack{N}}\frac{\sin\de{\theta_i-\theta_j}}{r_{ij}^\alpha} \,,
\label{eq:Hamilton_Eqs}
\end{equation}
where $r_{ij}=|\mathbf{r}_i-\mathbf{r}_j|$ measures the distance 
between rotators at sites~$i$ and~$j$ in lattice units, and it is defined as the minimal
one, given that periodic conditions will be considered. 
The parameter~$\alpha \ge 0$ controls the interaction range, whereas 
the scaling prefactor $1/\Ntil$  in the potential energy of
Hamiltonian~\eqref{eq:Hamiltonian_XY}
is introduced to make the energy extensive for all values of~$\alpha/d$, where~\cite{JundKimTsallisPRB1995, TsallisFractals1995, 
AnteneodoTsallisPRL1998, TamaritAnteneodoPRL2000, 
CampaGiansantiMoroniJPA2003}
\begin{equation}
\Ntil = \df{1}{N}\sum_{\substack{i=1}}^{\substack{N}} \sum_{\substack{j\neq i}}^{\substack{N}} \df{1}{r_{ij}^{\alpha}}
= \sum_{\substack{j= 2}}^{\substack{N}} \df{1}{r_{1j}^{\alpha}}
\hspace{0.5cm}
(\Ntil =2d \,\,{\rm for}\,\, \alpha\to\infty) \,,
\label{eq:Ntil}
\end{equation}
the $0 \le \alpha/d \le 1$ ($\alpha/d>1$) regime being hereafter referred to as long-range (short-range).
{Notice that when $\alpha=0$, we get $\Ntil=N-1\sim N$, so that Eq.~\eqref{eq:Hamiltonian_XY} recovers the HMF model.}

The results that follow were obtained from microcanonical molecular-dynamical 
simulations of a single realization of the system defined in~\eqref{eq:Hamiltonian_XY}, 
considering fixed values for the number of rotators~$N$ and energy
per particle~$u$, so that the total energy~$E=Nu$ is a constant.
To integrate the~$2N$ equations of motion in~\eqref{eq:Hamilton_Eqs},  
we have used the Yoshida $4th$-order symplectic 
algorithm~\cite{YoshidaPLA1990YoshidaCMDA1993},
choosing an integration step in such a way to yield a conservation of 
the total energy within a relative fluctuation always smaller than~$10^{-5}$.
At the initial time, all rotators were started with~$\theta_i=0 \,(\forall i)$; moreover, 
each momentum~$p_i$ was drawn from a symmetric uniform distribution $p_i\in[-1,1]$,
and then rescaled to achieve the desired energy~$u$,  
as well as zero total angular momentum $P=\sum_i p_i=0$, which also is 
a constant of motion.
As verified by many authors (see, e.g., Refs.~\cite{EttoumiFirpoPRE2013, BachelardKastnerPRL2013, CirtoAssisTsallisPA2014}), 
the model in Eq.~\eqref{eq:Hamiltonian_XY} exhibits a QSS 
for $0 \le \alpha/d<1$ and $u \simeq 0.69$, after which, a crossover to a
state whose temperature coincides with the one obtained within BG
statistical mechanics~\cite{AntoniRuffoPRE1995, 
CampaGiansantiMoroniJPA2003} occurs; herein 
we explore further properties of this model for the energy $u=0.69$.

\begin{figure*}
\centering
\subfigure[Momentum]{
    \includegraphics[width=0.80\linewidth]{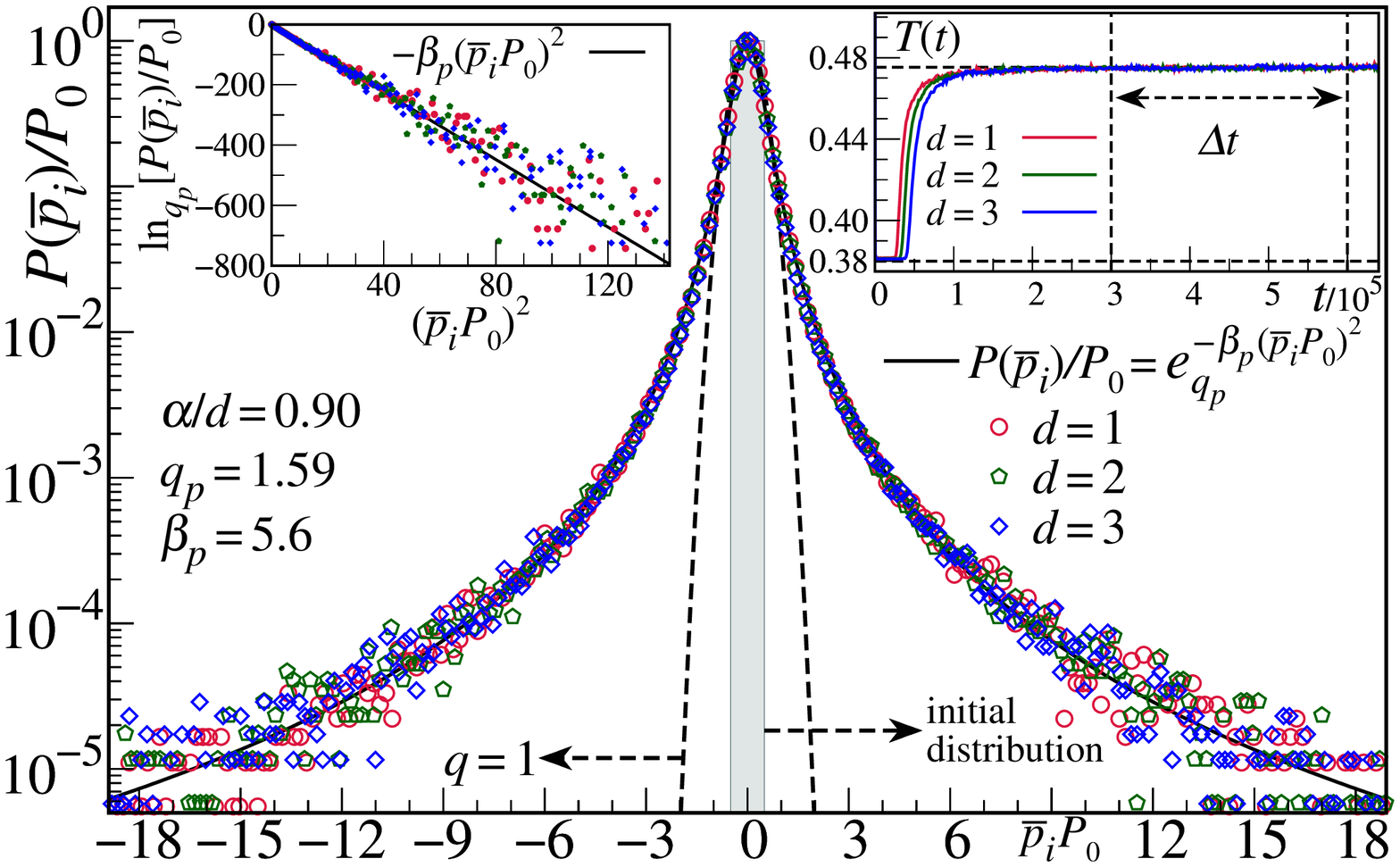}
  }
\subfigure[Energy]{
    \includegraphics[width=0.80\linewidth]{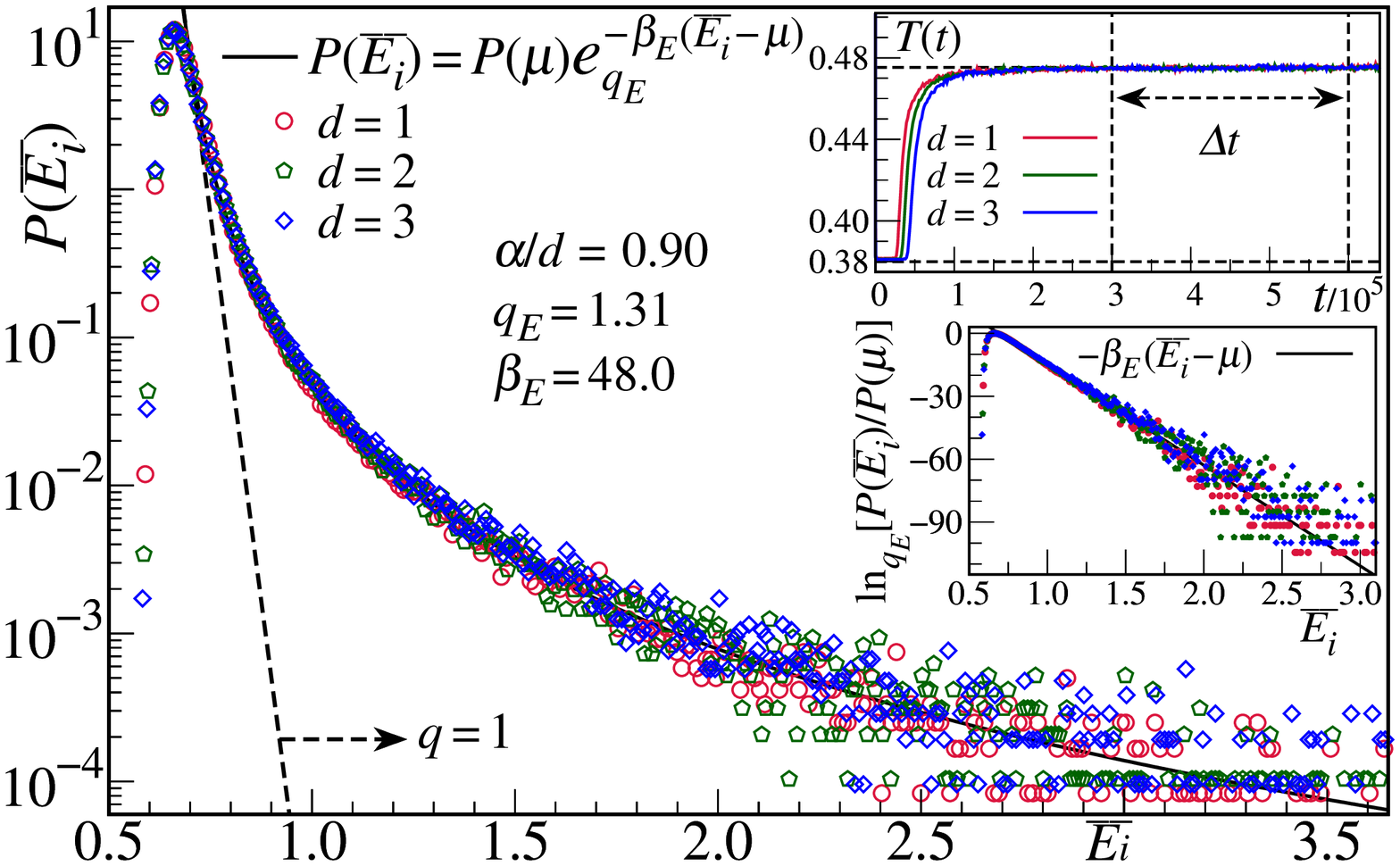}
  }
\protect\caption{\textsf{%
Distributions of time-averaged momenta $\bar{p}_i$ and 
energies $\bar{E}_i$ (with $\tau=1$) 
for $\alpha/d=0.9$, in $d=1$,$2$ and $3$ dimensions.   
The simulations were carried for the energy per particle $u=0.69$ and total number of rotators $N=1000000$.
\textbf{(a)} Distribution $P(\bar{p}_i)$ is shown  [$P_{0} \equiv P(\bar{p}_i=0)$]; the full line is a $q$-Gaussian 
with~$q_{p}=1.59$ and~$\beta_{p}=5.6$; the dashed 
line is a Gaussian $(q=1)$. 
The left inset shows the same data in a $q$-logarithm {\it versus} squared-momentum
representation; a straight line is obtained as expected (since $\ln_q(e_q^{x})= x$).
\textbf{(a)} The full line represents the $q$-exponential 
$P(\bar{E}_i)=P(\mu)\exp_{q_{E}}[-\beta_{E}(\bar{E}_i-\mu)]$, with 
$q_{E}=1.31$ ($\beta_{E}=48.0$, $\mu=0.69$, and $P(\mu)=12$); 
the corresponding exponential (dashed line) is also shown for comparison. 
Since the density of states is necessary to reproduce the entire range of data, 
the parameter $\mu$ was introduced in the fitting.
The bottom inset shows a straight line by using the $q$-logarithm in the ordinate.
The kinetic temperature~$T(t) \equiv 2K(t)/N$, and time window~$\Delta t$ 
along which the time averages were calculated, coincide in both cases
(shown as insets). 
In all plots one notices the collapse of all dimensions with nearly the same 
value of~$q$.  
}}
\label{Fig:figure1}
\end{figure*}

In Fig.~\ref{Fig:figure1} we present results for distributions of 
momenta [Fig~\ref{Fig:figure1}(a)]
and energies [Fig~\ref{Fig:figure1}(b)] of the model on 
hypercubic lattices ($d=1,2$ and 3), in the 
long-range regime, more specifically, $\alpha/d=0.9$.
In the insets on the right-hand sides
we show the time evolution of the kinetic temperature~$T(t)=2K(t)/N$,
as well as the time interval considered for the histograms.  
These distributions were calculated by 
registering $n$ times, e.g., the momenta~$p_i(t)$ ($\forall i$), 
at successive times separated by an interval~$\tau$, and then, 
following the Central Limit Theorem recipe, the 
arithmetic average $\bar{p}_i=\tfrac{1}{n}\sum_{k=0}^{n-1}p_i(t_0+k\tau)$ 
was obtained, leading to a histogram of these $N$ arithmetic averages.
Notice that such a recipe yields precisely a time average in this case
(associated with the time window $\Delta t = n\tau$), a situation that frequently 
corresponds to real experiments.
In order to improve the statistics of the histograms, we have considered  
rather large systems, up to $N=10^6$,
for a single numerical realization in each dimension~$d$. 
Curiously, although the kinetic temperature coincides with the BG prediction, 
the momentum distribution is quite distinct from a Gaussian; indeed, due to the
long-range nature of the interactions, the resulting distribution exhibits
a $q$-Gaussian form, as already observed in previous~$(d,\alpha)=(1,0)$
works~\cite{PluchinoRapisardaTsallisEPL2007, PluchinoRapisardaTsallisPA2008,%
CirtoAssisTsallisPA2014}; Fig~\ref{Fig:figure1}(a) extends these investigations to higher dimensions. 
Repeating the foregoing procedure to the 
one-particle energies~$E_i(t)$ [cf. Eq.~\eqref{eq:Hamiltonian_XY}],
the distributions in Fig~\ref{Fig:figure1}(b) are obtained.
One sees that a $q$-Gaussian emerges as the distribution of momenta (instead 
of a  Maxwellian), whereas a $q$-exponential appears for the energies (instead of 
the exponential Boltzmann weight). 
The results of Figs.~\ref{Fig:figure1}(a) 
and~\ref{Fig:figure1}(b)
are clearly out of the BG world and in close agreement
with the predictions of $q$-generalized statistical mechanics.

\begin{figure*}
\centering
\subfigure[]{
    \includegraphics[width=0.75\linewidth]{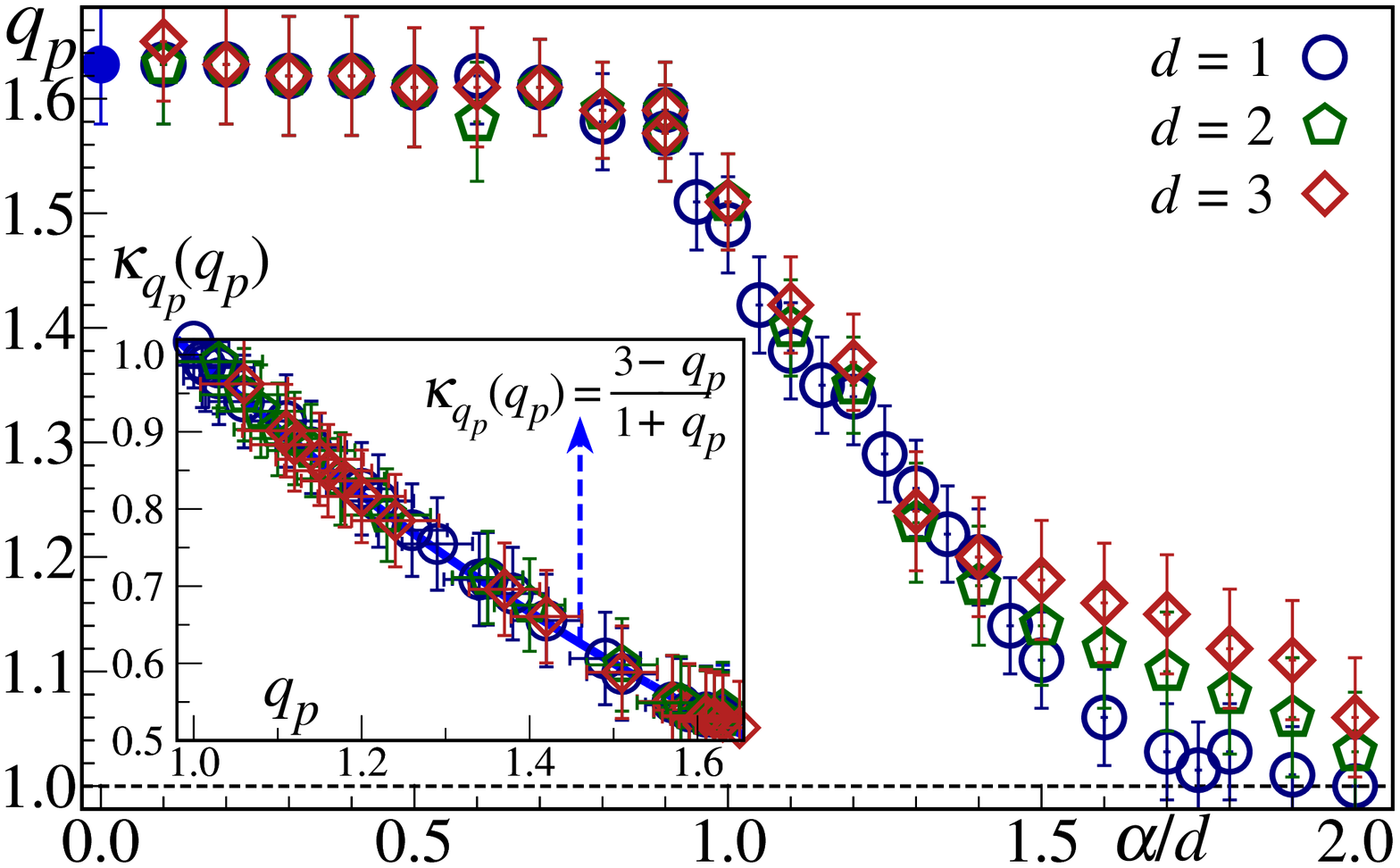}
  }
\subfigure[]{
    \includegraphics[width=0.75\linewidth]{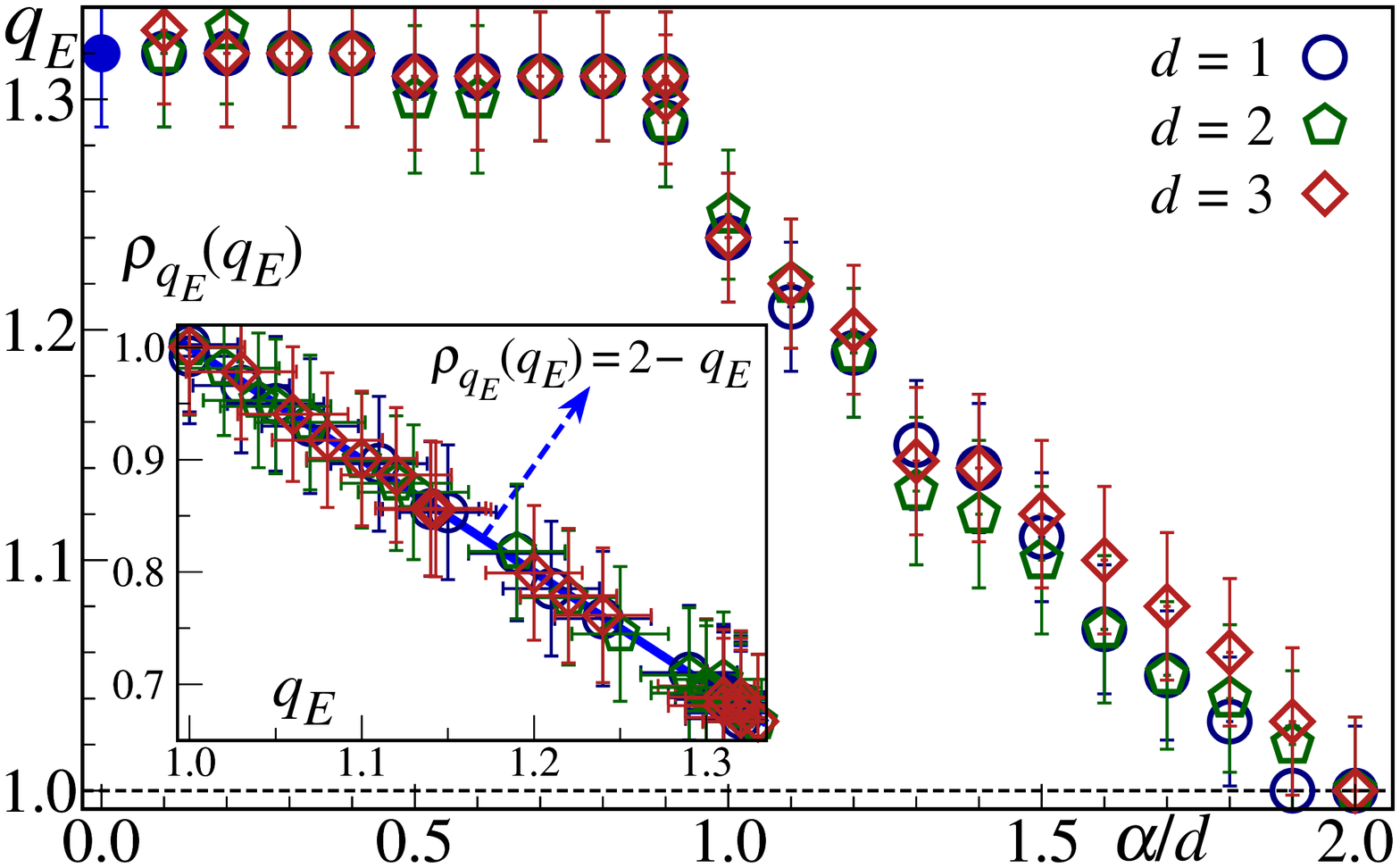}
  }
\protect\caption{\textsf{
$(\alpha/d)$-dependence of the indices $q_{p}$ and $q_{E}$ associated respectively with the distributions of time-averaged 
momenta $\bar{p}_i$ and energies $\bar{E}_i$  for $d=1,2,3$ and $u=0.69$. The insets show
the corresponding $q$-kurtosis~(a) and $q$-ratio~(b), compared to the 
analytical results (solid curves, see Eqs.~\eqref{eq:qKurtosis} and~\eqref{eq:qRatio}). 
The full bullets correspond to the values for $\alpha=0$.  
Notice that, within the error bars, the indices $q$ remain constant
for $0 \leq \alpha/d \leq 1$, and approach unit
only around $\alpha/d=2$ (see text). 
}}
\label{Fig:figure2}
\end{figure*}

The histograms shown in 
Figs.~\ref{Fig:figure1}(a)-(b)
were obtained from numerical simulations of~$N=10^6$ rotators with $\tau=1$ (5 integration steps) and $n=300000$; 
however, a systematic study for several values of~$\alpha/d$ was carried 
by considering smaller number of 
rotators (and consequently, different time windows), 
due to computational costs. 
In Figs.~\ref{Fig:figure2}(a)-(b)~we present the values 
of $q$ obtained from the distributions of time-averaged 
momenta $\bar{p}_i$ and energies $\bar{E}_i$ (labeled by 
$q_{p}$ and $q_{E}$, respectively) versus
$\alpha/d$ in $d=1$,$2$ and $3$ dimensions.
It should be mentioned that the results of 
Fig.~\ref{Fig:figure2}(a) are in good 
agreement with previous studies of the 
$d=1$ case~\cite{CirtoAssisTsallisPA2014}; here we also investigate $d=2,3$.  
A remarkable collapse is shown (within error bars) as a function of $\alpha/d$ for all dimensions; 
notice also that, generically, $q_{p}$ and $q_{E}$ do not coincide.
Intriguingly, these values of $q$ do 
not attain unit around $\alpha/d=1$, but rather at some
higher value, close to
$\alpha/d =2$.
This fact has also been observed in recent simulations of other models
with power-law decay of interactions:
(i) a one-dimensional quantum Ising 
ferromagnet~\cite{HaukeTagliacozzoPRL2013};
(ii) a Fermi-Pasta-Ulam-like one-dimensional Hamiltonian with a quartic coupling constant decaying with the distance between
oscillators~\cite{ChristodoulidiTsallisBountisEPL2014,ChristodoulidiBountisTsallisDrossosJSTAT2016,BagchiTsallisPA2018};
(iii) scale-free complex networks~\cite{BritoSilvaTsallisSR2016,NunesBritoSilvaTsallisJSTAT2017}.
Similarly to the present investigation, in these previous works
three distinct regimes were found,
namely, a non-BG long-range interacting regime $(0 \leq \alpha/d \leq 1)$,
a non-BG short-range one $(1 < \alpha/d \leq a_{c})$, and the standard BG short-range
regime ($\alpha/d > a_{c}$); for some classical Hamiltonians, $a_{c} \approx 2$, whereas, for complex networks, $a_{c} \approx 5$.  
The existence of these three regimes might be related to ergodicity and phase-space structure. More precisely, strong indications exist that, for $0 \leq \alpha/d \leq1$ ($\alpha/d >1$), weak (strong) chaos emerges~\cite{AnteneodoTsallisPRL1998,BagchiTsallisPRE2016}. How come an intermediate region ($1<\alpha/d<a_c$) exists, which is ergodic and nevertheless non-BG? A plausible explanation is that, similarly to 
the web map~\cite{RuizTirnakliBorgesTsallisPRE2017},
ergodicity takes place in a multifractal-like region and not in the entire phase space (or in a nonzero Lebesgue measure of it).
Furthermore, it should be mentioned that 
the results  $q_{p} \neq q_{E}$ could be due to 
finite-size effects, but this point deserves further investigation. 
Indeed, the plethoric results pointing out in many systems the existence of $q$-triplets and related 
structures~\cite{TsallisGellMannSatoPNAS2005,TsallisEPJST2017} 
could in principle emerge here as well, thus leading to values of $q$'s that differ among them for different 
basic quantities. These values could satisfy relations among them which would leave only a small 
number as independent ones, being all the others functions of those few.

To check the $q$-Gaussian fits in the one-particle momentum distribution, 
we used the $q$-kurtosis~\cite{CirtoAssisTsallisPA2014}, 
\begin{equation}
\kappa_q\de{q} \equiv \df{1}{3}\df{\media{p^4}_{2q-1} }{\media{p^2}_q^2}
= \frac{3-q}{1+q}~, 
\label{eq:qKurtosis}
\end{equation}
whereas for the energies, we used the
$q$-ratio~\cite{BagchiTsallisPA2018}
\begin{equation}
\rho_q\de{q} \equiv \df{1}{2}\df{\media{\epsilon^2}_{2q-1} }{\media{\epsilon}_q^2} = 2-q~,   
\label{eq:qRatio}
\end{equation}
the $q$-moments being defined 
as~\cite{TsallisMendesPlastinoPA1998, TsallisLivro2009, CirtoAssisTsallisPA2014} 
\begin{equation}
\media{x^{m}}_{f(q)}=\frac{\int\!\rd x \,x^{m} [P(x)]^{f(q)}}{\int\! \rd x[P(x)]^{f(q)}}\;
\hspace{0.5cm}[f(q) \equiv 1+m(q-1)]     \,,  
\label{eq:qmoments}
\end{equation}
with $x=p^{2}$ ($x=\epsilon$) for Eq.~\eqref{eq:qKurtosis} [Eq.~\eqref{eq:qRatio}]. 
From the momentum and energy histograms we have 
computed~$\kappa_{q_p}(q_p)$  
and~$\rho_{q_E}(q_E)$ for several values of~$\alpha/d$: see insets of 
Figs.~\ref{Fig:figure2}(a)-(b). 
These numerical data exhibit good agreement with the above analytical results. 
Naturally, neither numerical nor experimental results will ever produce mathematical proofs of 
whatever analytical expressions of any theory. Interesting illustrations of this trivial fact have been discussed 
some years ago for compact-support numerical 
distributions~\cite{HilhorstSchehrJSTAT2007,RodriguezSchwammleTsallisJSTAT2008}.  
Let us however emphasize that, in the present case, we are focusing on fat-tailed distributions on which 
such numerical coincidences certainly are much harder to occur along many decades.

\begin{figure}
\centering
  \includegraphics[width=0.80\linewidth]{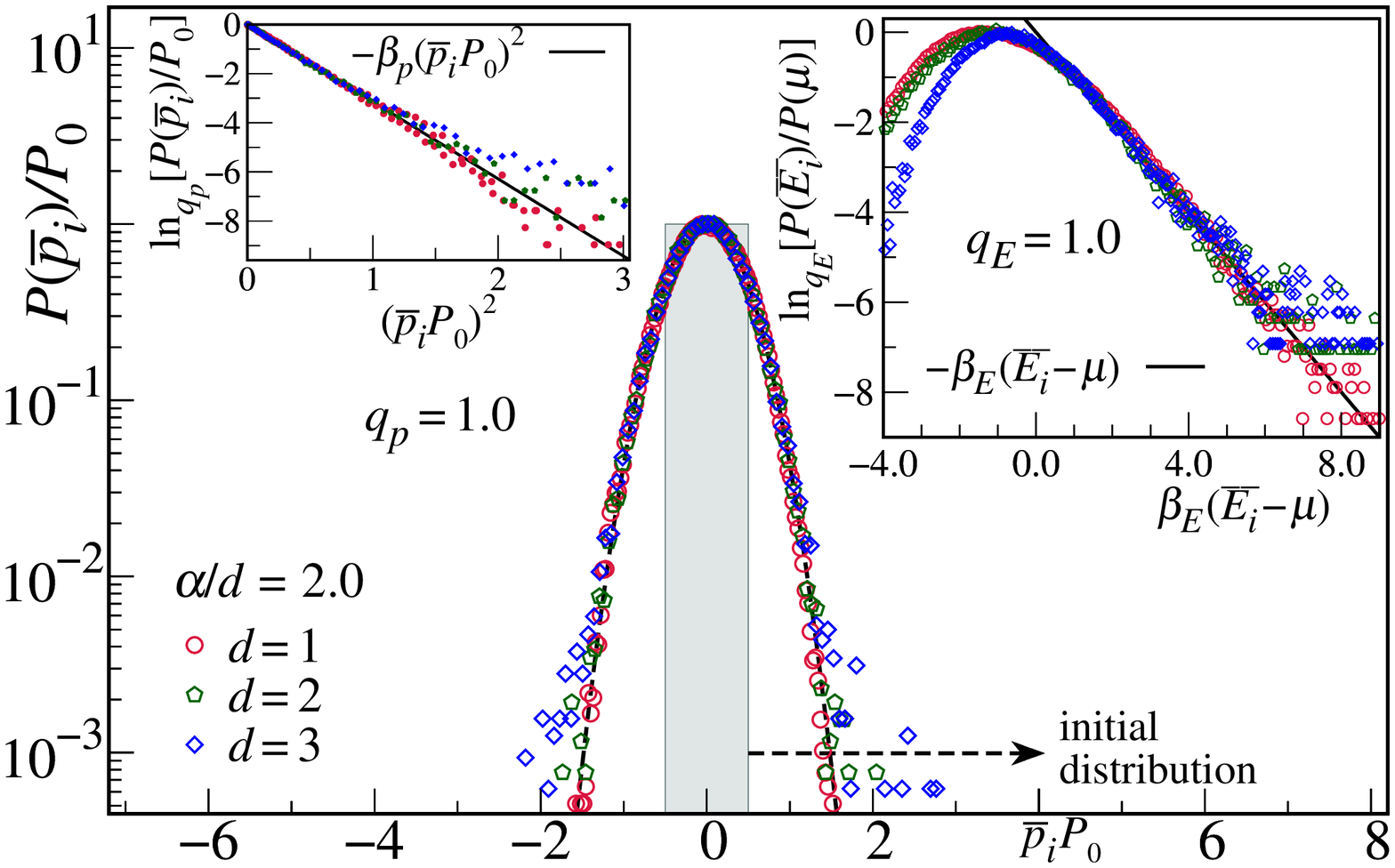}
\protect\caption{\textsf{Distributions for time-averaged momenta $\bar{p}_i$ and 
energies $\bar{E}_i$ are shown 
for $\alpha/d=2$  ($d=1,2,3$).
For the momenta we have used conveniently scaled variables
[like in Fig~\ref{Fig:figure1}(a)], and the full line is the Maxwellian 
$(q=1)$; the left inset shows the same data in a 
logarithm versus squared-momentum representation.
The right inset exhibits $\ln[P(\bar{E}_i)/P(\mu)]$
{\it versus} $\beta_{E}(\bar{E}_i-\mu)$. Similarly to  
Fig~\ref{Fig:figure1}(b), we verify the appearance of $d$-dependent 
densities of states; 
the full line is an exponential in the 
variable $\beta_{E}(\bar{E}_i-\mu)$. 
}}
\label{Fig:figure3}
\end{figure}

For completeness, in Fig.~\ref{Fig:figure3} we 
present distributions for time-averaged momenta $\bar{p}_i$ and 
energies~$\bar{E}_i$, for the short-range-interaction 
regime $\alpha/d=2$ ($d=1,2,3$).
For the momenta, our data are well fitted by a Gaussian, 
whereas for the energies one notices a straight line for 
$\bar{E}_i > \mu$ (see right inset), landmark of the Boltzmann
weight. 
For $\alpha/d=2$, the lattice dimensionality starts playing an 
important role, clearly detected in the numerical simulations:
(i) Different dimensions are characterized by distinct 
density of states, so that the collapse of the energy
distributions only occurs for $\bar{E}_i > \mu$; 
(ii) The number of nearest-neighbor rotators increases
with $d$, which directly reflects on the computational time
of the simulations (for this reason, we have considered 
$N=262144$ for $d=1$, and 
$N=46656$ for $d=2,3$).

\begin{figure}
\centering
\includegraphics[width=0.75\linewidth]{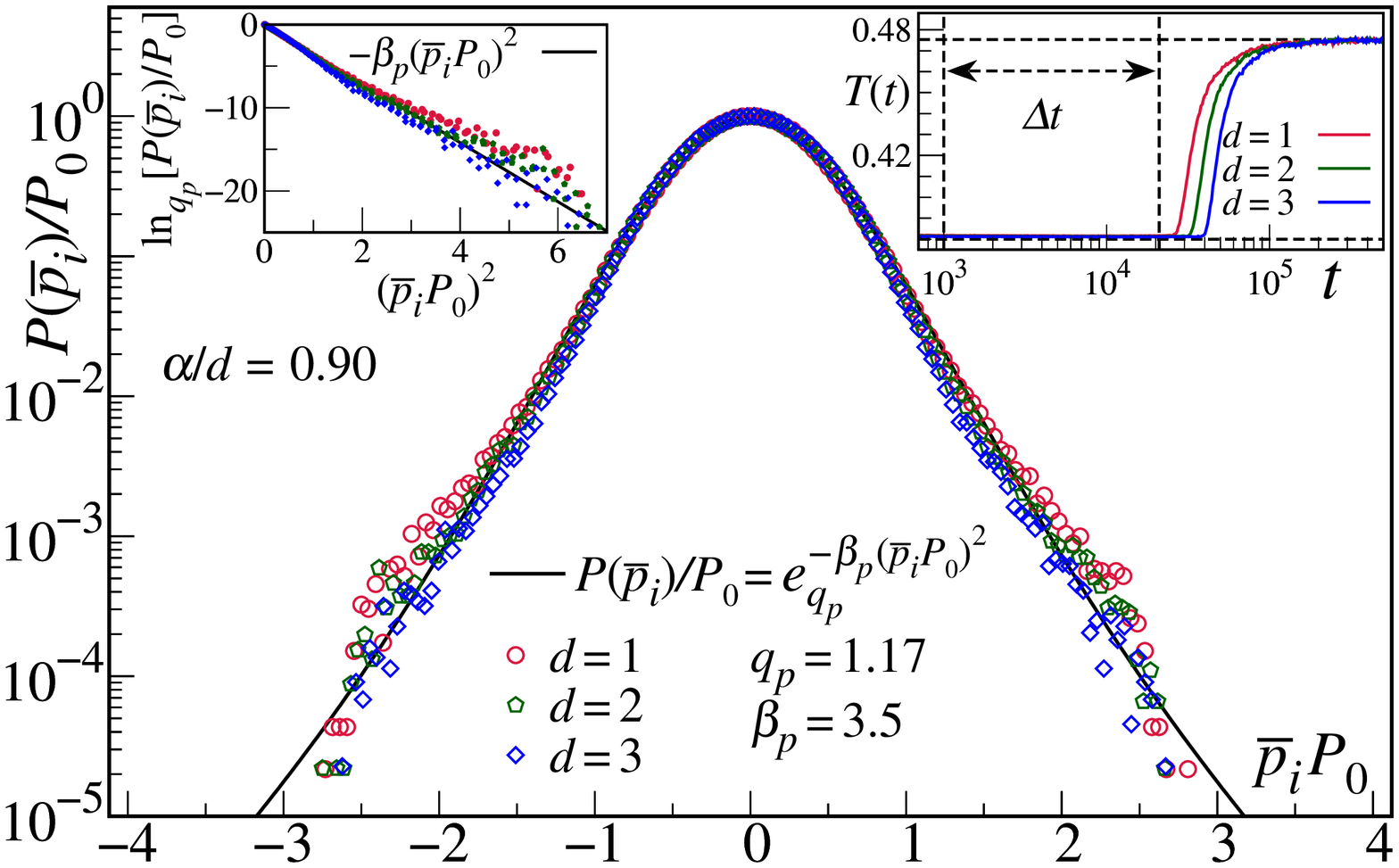}   
\caption{\textsf{Distributions of time-averaged momenta $\bar{p}_i$ for  
the same parameters used in Fig.~\ref{Fig:figure1}(a), 
but for a different time window. The left inset presents the same data in a $q$-logarithm versus 
squared-momentum
representation; the full straight line is a $q$-Gaussian with~$q=1.17$.
One should notice the collapse of all dimensions with nearly the same value of~$q$.
The right inset shows~$T(t) \equiv 2K(t)/N$ and the time-window~$\Delta t$ 
along which the time averages were calculated ($t\in[1,21]\times 10^3$).  
}}
\label{Fig:figure4}
\end{figure}

In Fig.~\ref{Fig:figure4} we present distributions of momenta for the same 
systems considered in Fig.~\ref{Fig:figure1}(a), but now for a substantially earlier time 
window ($t\in[1,21]\times 10^3$), within the QSS. Once again, the collapse of all
three histograms ($d=1,2,3$) into a single $q$-Gaussian is observed, although for a smaller value, $q_{p}=1.17$.
The duration $t_\text{QSS}$ of such a QSS increases 
with $N$ and decreases with $\alpha/d$~\cite{CampaDauxoisRuffoPR2009, BachelardKastnerPRL2013, PluchinoRapisardaTsallisPA2008, CirtoAssisTsallisPA2014};
we verified that $t_\text{QSS} \sim N^{\gamma(\alpha/d)}$ with $\gamma(0.9) \simeq 0.6$.
Therefore, the analysis of histograms in the QSS must take into account the window $\Delta t$ for the time averages, according to $N$ and $\alpha/d$. A detailed study of these effects is out of our present scope and represents a matter for future investigations. 
 
To summarize, we have presented molecular-dynamics results for 
a classical inertial XY model, on $d$-dimensional lattices ($d=1,2,3$),
characterized by interactions with a variable range.
These interactions decay with the distance $r_{ij}$ between rotators at sites $i$ and
$j$, like~$1/r_{ij}^{\alpha}$ ($\alpha \geq 0$), 
so that, by increasing gradually the parameter~$\alpha$, one interpolates
between the infinite-range-interaction ($\alpha=0$) and the nearest-neighbor
($\alpha\to\infty$) limits.  
Our numerical analyses strongly suggest that crucial properties, like 
probability distributions, 
depend on the ratio $\alpha/d$, rather than on $\alpha$ and $d$ separately. 
For sufficiently high values of $\alpha/d$ we have found
Maxwellians for the momenta, as well as the Boltzmann weight for the energies. 
On the other hand, in the long-range-interaction 
regime ($\alpha/d < 1$), we have observed
$q$-Gaussians for the time-averaged momenta, 
as well as $q$-exponential distributions for the time-averaged energies,
thus undoubtedly falling out of the scope of Boltzmann-Gibbs statistical mechanics. 
The present study corroborates investigations on different long-range systems, 
such as the extended Fermi-Pasta-Ulam model and complex networks, thus showing 
that central properties depend on the 
ratio  $\alpha/d$. In particular, the values of the indices $q$ herein found vary
as $q=q(\alpha/d)$, in full agreement with the theoretical 
expectations of nonextensive statistical mechanics. 

\begin{acknowledgments}
{We acknowledge useful conversations with D. Bagchi, E.P. Borges, T. Bountis, H. Christodoulidi, E.M.F. Curado, G. Ruiz, 
A.M.C. Souza, and U. Tirnakli. We benefited from partial financial support by CNPq, 
FAPERJ, and CAPES (Brazilian agencies), and the John Templeton Foundation (USA).}
\end{acknowledgments}
\clearpage

\end{document}